\documentclass[aps,pre,showpacs,reprint,groupedaddress]{revtex4-1}
\usepackage{amsmath}
\usepackage{lmodern}
\usepackage{graphicx}
\usepackage{amssymb} 
\pdfoutput=1
\usepackage{physics}
\usepackage{color}
\usepackage[colorlinks,citecolor=blue,breaklinks=true]{hyperref}

\begin{document}
\title{Entrainment of a network of interacting neurons with minimum stimulating charge}

\author{Kestutis~Pyragas}
 \email{kestutis.pyragas@ftmc.lt}
 \affiliation{Center for Physical Sciences and Technology, LT-10257 Vilnius, Lithuania}

\author{Augustinas~P.~Fedaravi\v{c}ius}
 \affiliation{Center for Physical Sciences and Technology, LT-10257 Vilnius, Lithuania}
 
\author{Tatjana~Pyragien\.{e}}
 \affiliation{Center for Physical Sciences and Technology, LT-10257 Vilnius, Lithuania}

\author{Peter~A.~Tass}
 \affiliation{Department of Neurosurgery, Stanford University, Stanford, CA, United States}

\begin{abstract}

Periodic pulse train stimulation is generically used to study the function of the nervous system and to counteract disease-related neuronal activity, e.g., collective periodic neuronal oscillations. The efficient control of neuronal dynamics without compromising brain tissue is key to research and clinical purposes. We here adapt the 
minimum charge control theory, recently developed for a single neuron, 
to a network of interacting neurons exhibiting collective periodic oscillations.  We present a general expression for the optimal waveform, which provides an entrainment of a neural network to the stimulation frequency with a minimum absolute value of the stimulating current. As in the case of a single neuron, the optimal waveform is of bang-off-bang type, but its parameters are now determined by the parameters of the effective phase response curve of the entire network, rather than of a single neuron. The theoretical results are confirmed by three specific examples: two small-scale networks of FitzHugh-Nagumo neurons with synaptic and electric couplings, as well as a large-scale network of synaptically coupled quadratic integrate-and-fire neurons.
\end{abstract}


\maketitle

\section{Introduction}
\label{sec:introduction}

The synchronization of coupled dynamical elements is of great interest to the physical, chemical, and biological sciences~\cite{Kuramoto2003,winf01,pikov01,izhi07}. In the nervous system, synchronization processes play an important role, as they are responsible for information processing and motor control. However, pathological, excessive synchronization can severely impair
brain function and is characteristic of several neurological disorders~\cite{Hammond2007,Uhlhaas2010}. Thus, the control of synchronization processes in neural systems is a demanding clinical problem. Over the past three decades, several control methods have been developed and applied. High-frequency ($>100$ Hz) deep brain stimulation (DBS)  \cite{Benabid1991,Krack2003,Deuschl2006,kring07} 
is an established and powerful therapeutic tool for the treatment of patients with Parkinson's disease, essential tremor, dystonia and even psychiatric disorders~ 
\cite{perlmutter2006deep,
koller2001long,vidailhet2005bilateral,
hardesty2007deep,krack2010deep}. 
Conventional DBS has only acute effects, i.e. neither clinical \cite{Temperli2003} nor electrophysiological \cite{Kuhn2008} effects persist after switching off conventional DBS.

The computationally developed method of coordinated reset (CR)-DBS
\cite{tass2003,Tass2006} is characterized by long-lasting, sustained effects, which persist after cessation of stimulation 
\cite{tass2012,Wang2016,Adamchic2014}. Standard DBS, CR-DBS as well as theta burst DBS, i.e. the delivery of periodic sequences of electrical bursts, recently tested in a short-term trial~\cite{Horn2020}, employ periodic pulse train stimulation. For all of these approaches, it is desirable to achieve a therapeutic effect with minimal interference with nerve tissue. To avoid side effects, it is crucial to achieve therapeutic effects with minimal stimulation current~\cite{Lozano2019,Feng2007,Wilson2020}. This raises the problem of finding the optimal waveform for stimulation.

In the field of theoretical and computational neuroscience, the problem of optimal synchronization is usually formulated as a control with minimal energy~ \cite{Moehlis2006,Kiss2010,Dasanayake2011,Nabi2013,nabi2013minimum,Li2013,dasanayake2014,dasanayake2015constrained,Pyragas2015,Wongsarnpigoon2010}. The goal is to obtain the optimal waveform of a periodic stimulation current to entrain a given spiking neuron with minimal stimulation energy. Another approach for optimal synctronization was developed in Refs.~\cite{Tanaka2014,tannaka2014JPA,Tanaka2015}. The authors introduced a general form of a functional and considered the problem of maximizing the width of the entrainable frequency detuning (the width of the Arnold tongue) of a spiking neuron for the fixed value of the functional. This approach can be applied to an ensemble of non-interacting neurons with distributed frequencies. 

The minimum-energy control strategies reduce the energy consumption of an implantable pulse generator but do not guarantee minimal damage of the neural tissue. Recently, we proposed an alternative, minimum-charge control strategy, which aims at reducing damage to neural tissue~\cite{Pyragas2018}. Neurological stimulation protocols typically use periodic, charge-balanced, biphasic stimuli, usually asymmetrical in shape~\cite{Coffey2009}. Typically, the first phase of the stimulus depolarizes the cell membrane and the second pulse brings the net charge balance in the electrode back to zero~\cite{Hofman2011}. One of the important factors on which the threshold for tissue damage depends is the charge per phase of a stimulus pulse~\cite{McCreery1990,Shannon1992}. The magnitude of charge is defined by the product of the amplitude and width of the pulse. For the periodic charge-balanced stimulation, this quantity is proportional to the mean absolute value of the stimulating current. The latter was chosen as a performance measure in our control algorithm~\cite{Pyragas2018} in order to minimize the integral charge transferred to the neuron in both directions  during the stimulation period. The consideration in Ref.~\cite{Pyragas2018} was limited to a single neuron. In this paper, we apply this approach to a network of interacting neurons exhibiting collective periodic oscillations. We use the results of a recently developed phase reduction theory for arbitrary networks of coupled heterogeneous dynamical elements~\cite{Nakao2018}. We also apply our approach to a large-scale heterogeneous network of globally coupled quadratic integrate-and-fire (QIF) neurons, which can be reduced to an exact low-dimensional macroscopic model in the infinite-size limit~\cite{Montbrio2015,Ratas2016}. 

The paper is organized as follows. In Sec.~\ref{sec:problem} we formulate the  problem, and in Sec.~\ref{sec:Optimaltheory} we give a general expression for the optimal waveform, which ensures the minimum charge entrainment of any network of interacting neurons to the frequency of stimulating current. This general theoretical result is then numerically demonstrated for small-scale networks of synaptically and electrically coupled FitzHugh-Nagumo (FHN) neurons as well as for a large-scale network of QIF neurons in the Sec.~\ref{sec:numerics}. We finish the paper with a discussion and conclusions presented in the Sec.~\ref{sec:conclusions}

\section{Problem formulation}
\label{sec:problem}
We consider a general heterogeneous network of $N$ coupled Hodgkin-Huxley-type neurons under periodic stimulation
\begin{subequations}
\label{eq:gen_neuron}
\begin{eqnarray}
\dot{v}_i &=& F_i(v_i,\mathbf{w}_i)+\sum_{j=1}^N H_{ij}(v_i,v_j)+I_i(\omega t),  \label{eq:gen_neurona}\\
\dot{\mathbf{w}_i} &=& \mathbf{G}_i (v_i,\mathbf{w}_i) \quad (i=1,\ldots, N).  \label{eq:gen_neuronb}
\end{eqnarray}
\end{subequations}
Here the scalar $v_i$ and the vector $\mathbf{w}_i \in \mathbb{R}^{n}$ are the membrane potential and the recovery variable of the $i$th neuron, respectively. The function $F_i(v_i,\mathbf{w}_i)$ describes the sum of currents flowing through the ion channels of the $i$th neuron and the function $H_{ij}(v_i,v_j)$ defines the coupling between the $i$th and $j$th neuron. $I_i(\omega t)$ is a periodic stimulating current applied to the $i$th neuron. It satisfies $I(\omega t+2\pi)=I(\omega t)$, where $\omega$ is the stimulation frequency and $T=2\pi/\omega$ is the period of stimulation. Equation~(\ref{eq:gen_neuronb}) describes the dynamics of the recovery variable $\mathbf{w}_i$, where the function $\mathbf{G}_i (v_i,\mathbf{w}_i)$ represents the ionic channel dynamics. The dimension $n$ of the vector variable $\mathbf{w}_i$ as well as the functions $F_i$ and $\mathbf{G}_i$ are defined by the specific neuron model.
 
We assume that without stimulation [$I_i(\omega t)=0$] the entire network exhibits stable collective limit cycle oscillations with a period $T_0$ and a frequency $\omega_0=2\pi/T_0$. Our goal is to find the optimal waveform for the stimulating currents $I_i(\omega t)$, which ensures the entrainment of the network to the stimulation frequency $\omega$ with minimum integral charge transferred to neurons in both directions during the stimulation period. For a single neuron, such a problem was considered in our recent publication~\cite{Pyragas2018}. Below we will show that, under certain assumptions, the results of Ref.~\cite{Pyragas2018} can be adapted to a network of interacting neurons.

For sufficiently small stimulating currents $I_i(\omega t)$, the phase reduction method \cite{Kuramoto2003,izhi07,nakao2016phase} can be applied to reduce Eqs.~\eqref{eq:gen_neuron} to a single scalar phase equation. Our equations represent a particular case of equations considered in a recent publication~\cite{Nakao2018} for which such a reduction has been performed, and thus we can directly use these results to write down the reduced equations for our problem. The dynamics of the $(n+1)N$ dimensional system of ordinary differential Eqs.~\eqref{eq:gen_neuron} can be approximated by phase equation 
\begin{equation}
	\dot{\vartheta} = \omega_0 + \sum_{i=1}^N z_i(\vartheta) I_i(\omega t)	\label{eq:phase}
\end{equation}
for the collective phase $\vartheta(t)$. Here $z_i(\vartheta)$ is the $2\pi$-periodic phase response curve (PRC) of the $i$th neuron.

The PRCs are derived from a free [$I_{i}(\omega t)=0$] network model. It is convenient to introduce $(n+1)$-dimensional vectors 
\begin{equation}
\mathbf{X}_i=\left(
\begin{array}{c}
v_i\\
\mathbf{w}_i\\
\end{array}
\right),\
\mathbf{\Phi}_i(\mathbf{X}_i)=\left(
\begin{array}{c}
F_i(v_i,\mathbf{w}_i)\\
\mathbf{G}_i(v_i,\mathbf{w}_i)\\
\end{array}
\right)
	\label{eq:aux_vects}
\end{equation}
and rewrite the free network Eqs.~\eqref{eq:gen_neuron} in the form
\begin{equation}
\dot{\mathbf{X}}_i= \mathbf{\Phi}_i(\mathbf{X}_i)+\mathbf{k}\sum_{j=1}^N H_{ij}(v_i,v_j) \quad (i=1,\ldots, N), 
	\label{eq:free_netw}
\end{equation}
where $\mathbf{k}$ is a $(n+1)$ dimensional unity vector with the first component equal to one and all other components equal to zero. We denote the $T_0$ periodic stable limit-cycle  solution of the free network~\eqref{eq:free_netw} as 
\begin{equation}
\mathbf{X}_i^{(0)}(t)= \mathbf{X}_i^{(0)}(t+T_0) \quad (i=1,\ldots, N). 
	\label{eq:per_sol_free}
\end{equation}
Generally, the PRCs of the system~\eqref{eq:free_netw} are the $(n+1)$ dimensional $2\pi$-periodic vectors 
\begin{equation}
\mathbf{Q}_i(\vartheta)= [Q_i^{(1)},\ldots,Q_i^{(n+1)} ]^T \quad (i=1,\ldots, N). 
	\label{eq:PRC_vect}
\end{equation}
However, since the stimulating currents $I_i(\omega t)$ in the system~\eqref{eq:gen_neuron}  perturb only membrane potentials of neurons [Eq.~~\eqref{eq:gen_neurona}], their contribution to the phase dynamics in the Eq.~\eqref{eq:phase} is determined by the first components $Q_i^{(1)}(\vartheta)$ of the vectors  $\mathbf{Q}_{i}(\vartheta)$, which in Eq.~\eqref{eq:phase} are denoted as 
\begin{equation}
z_i(\vartheta) \equiv Q_i^{(1)}(\vartheta). 
	\label{eq:scalarPRC}
\end{equation}
Although we only need  the first components of the PRC vectors to find them, we have to solve the system of adjoint equations for the full PRCs vectors~\cite{Nakao2018}:
\begin{equation} \label{eq:adjoint}
\begin{split}
\omega_0 \frac{d}{d \vartheta}\mathbf{Q}_i(\vartheta)=& -A_i^T(\vartheta)\mathbf{Q}_i(\vartheta)-\mathbf{k}\sum_{j=1}^N M_{ij}(\vartheta) z_i(\vartheta)\\
&-\mathbf{k}\sum_{j=1}^N N_{ji}(\vartheta) z_j(\vartheta) \quad  (i=1,\ldots, N),
\end{split}
\end{equation}
where  $A_i(\vartheta)=\partial \mathbf{\Phi}_i(\mathbf{X}_i)/\partial \mathbf{X}_i$, $M_{ij}(\vartheta)=\partial H_{ij}(v_i,v_j)/\partial v_i$ and $N_{ij}(\vartheta)=\partial H_{ij}(v_i,v_j)/\partial v_j$ are Jacobian matrices of $\mathbf{\Phi}_i$ and partial derivatives of the scalar coupling functions $H_{ij}$ evaluated at $\mathbf{X}_i^{(0)}(\vartheta)=\mathbf{X}_i^{(0)}(\omega_0 t)$, respectively. The superscript $T$ denotes the matrix transpose. The PRCs have to satisfy the normalization condition
\begin{equation}
\sum_{i=1}^N \mathbf{Q}_j(\vartheta)\cdot \frac{d \mathbf{X}_i^{(0)}(\vartheta)}{d \vartheta}=1. 
	\label{eq:norm_cond}
\end{equation}
Thus, for a specific neural network model~\eqref{eq:gen_neuron}, PRCs $z_i(\vartheta)$ can be found by numerically solving the Eqs.~\eqref{eq:adjoint} for vectors $\mathbf{Q}_{i}(\vartheta)$  with the normalization condition~\eqref{eq:norm_cond}.

\section{Optimal waveform for entrainment of  a network of interacting neurons}
\label{sec:Optimaltheory}

In what follows, we assume that the subset 
$i \in \{i_k\} \equiv \{i_1,i_2,\ldots, i_M\}$ of  $M\le N$ network neurons are stimulated by the same 
 current $I(\omega t)$, while other neurons remain free from stimulation, i.e. we set
\begin{equation}
	I_i(\omega t) = 
	\begin{cases}
		I(\omega t) &\text{for} \ i \in \{i_1,i_2,\ldots, i_M\}\\*
		0, &\text{otherwise}
	\end{cases}
	\label{eq:current_distr}
\end{equation}
Then the phase Eq.~\eqref{eq:phase} simplifies to
\begin{equation}
	\dot{\vartheta} = \omega_0 + z(\vartheta) I(\omega t),	\label{eq:phase1}
\end{equation}
where 
\begin{equation}
z(\vartheta) =  \sum_{k=1}^M z_{i_k}(\vartheta).	\label{eq:eff_PRC}
\end{equation}
This assumption allows us to adapt the results of our recent publication~\cite{Pyragas2018}, where a minimum-charge control strategy for a single neuron was developed. The phase Eq.~\eqref{eq:phase1} has the same form as in the case of a single neuron, with the only difference that $z(\vartheta)$ is now the sum  of PRCs of stimulated neurons [Eq.~\eqref{eq:eff_PRC}] in a connected network, rather than the PRC of a single neuron. Thus, we can use the optimal waveform derived for the single neuron by replacing the PRC of a single neuron with an effective network's PRC defined by the Eq.~\eqref{eq:eff_PRC}.

Following Ref~\cite{Pyragas2018}, we briefly describe the minimum-charge control strategy for our network.  The aim of the strategy is to minimize the integral charge transferred to the neurons in both directions  during the stimulation period. We look for the optimal waveform $I(\omega t)=I^*(\omega t)$ that ensures an entrainment of a connected network~\eqref{eq:gen_neuron} to the frequency of stimulation with the minimum mean absolute value of the current injected into neurons:
\begin{equation}
	\mathcal{J}[I] = \frac{1}{T}\int_0^T \abs{ I(\omega t) } \dd{t}. 
	\label{eq:func}
\end{equation}
We minimize the functional \eqref{eq:func} with two additional, clinically relevant, conditions. We require that the stimulating current never exceeds some predefined minimal $I_-<0$ and maximal $I_+>0$ values, i.e.,
\begin{equation}
	 I_-\le I(\omega t)\le I_+
	\label{eq:bound_I} 
\end{equation}
for any time and introduce a clinically mandatory charge-balance condition 
\begin{equation}
	\int_0^T I(\omega t) \dd{t} = 0.
	\label{eq:charge_balance}
\end{equation}

For a sufficiently small frequency detuning
\begin{equation}
	\Delta \omega=\omega-\omega_0,
	\label{eq:mismatch}
\end{equation}
the optimization problem defined by Eqs.~\eqref{eq:phase1}, \eqref{eq:func}, 
\eqref{eq:bound_I}, and \eqref{eq:charge_balance} leads to the following optimal waveform~\cite{Pyragas2018}
\begin{equation}
	I^*(\vartheta) = I_+\Pi\qty( \frac{\vartheta}{\Delta\vartheta_+} ) + I_-\Pi\qty( \frac{\vartheta \pm \Delta \vartheta_{z}}{\Delta\vartheta_-} ), 
	\label{eq:opt_gen}
\end{equation}
which is a $2\pi$-periodic function $I^*(\vartheta)=I^*(\vartheta+2\pi)$ of the bang-off-bang type. Here $\Pi(x)$ is the rectangular function  that satisfies $\Pi(x) =1$ for  $\abs{x}<1/2$ and $\Pi(x) =0$ for  $\abs{x}>1/2$, and 
\begin{equation}
	\Delta \vartheta_{z}=\vartheta_{max} -\vartheta_{min} +\pi  \pmod{2\pi}-\pi
	\label{eq:delta_theta_z}
\end{equation}
is the phase difference between the location of the absolute maximum $\vartheta_{max}$ and the location of the absolute minimum $\vartheta_{min}$ of PRC \eqref{eq:eff_PRC} reduced to the interval $\Delta \vartheta_{z} \in[-\pi, \pi]$. The upper and lower signs in the argument of the second function in Eq.~\eqref{eq:opt_gen} correspond to $\Delta \omega>0$ and $\Delta \omega<0$, respectively. Thus, the waveform~\eqref{eq:opt_gen} consists of two rectangular pulses, one positive  located at $\vartheta=0$, and the other negative located at $\vartheta=-\Delta \vartheta_{z}$ for $\Delta \omega>0$ or at $\vartheta=\Delta \vartheta_{z}$ for $\Delta \omega>0$. 
The height of the positive (negative) pulse is $I_+$ ($I_-$) and its width is $\Delta\vartheta_+$ ($\Delta\vartheta_-$). The widths $\Delta\vartheta_{+,-}$ are defined as 
\begin{equation}
	\Delta\vartheta_{+,-} = \frac{2\pi\Delta\omega}{I_{+,-}( z_{max} - z_{min} )},
	\label{eq:widths}
\end{equation}
where $z_{max}=z(\vartheta_{max})$ and $z_{min}=z(\vartheta_{min})$ are the absolute maximum and absolute minimum of PRC \eqref{eq:eff_PRC}, respectively, therefore $z_{max}-z_{min}$ is the the amplitude of the PRC. The network \eqref{eq:gen_neuron} of interacting neurons oscillating with a frequency of $\omega_0$ is synchronized with the optimal stimulating current  $I^*(\omega t)$ of a frequency of $\omega$ at a minimum value of the functional \eqref{eq:func} equal to~\cite{Pyragas2018}:
\begin{equation}
	\mathcal{J}^* = \frac{2\abs{\Delta\omega}}{z_{max} - z_{min}}.
	\label{eq:optQs}
\end{equation}	

Summing up, in order to build the optimal waveform \eqref{eq:opt_gen} only limited information on the PRC \eqref{eq:eff_PRC} of the network is required, namely the phase difference $\Delta \vartheta_z$ between its absolute maximum and minimum,  and the amplitude $z_{max} - z_{min}$.
The optimal waveform $I^*(\vartheta)$ is of bang-off-bang type with periodically repeated positive and negative rectangular pulses separated by the distance $\Delta \vartheta_z$. The pulse sequence depends on the sign of the frequency detuning $\Delta \omega$. For $\Delta \omega>0$, this sequence is such that the positive pulse hits the network at the timing of the maximum of PRC while the negative pulse hits the network at the timing of the minimum of PRC. As a result, the maximum phase advancement occurs in the network during one period of the forcing. For $\Delta \omega<0$, the reverse pulse sequence is optimal.  The maximum phase delay of the network is achieved when the positive pulse hits the network at the minimum of the PRC, and the negative pulse hits the network at the maximum of the PRC. The amplitudes of the positive and negative pulses are equal to the permissible limits $I_+$ and $I_-$ of the stimulating current, respectively. The width of the positive $\Delta \vartheta_+$  (negative $\Delta \vartheta_-$) pulse is proportional to the frequency detuning $\Delta \omega$ and inversely proportional to the permissible limit $I_+$ ($I_-$).  It follows that the areas under the positive and negative pulses are equal to each other, so that the condition of charge-balance is satisfied.   

In the next section, we will numerically verify 
the general theoretical results derived 
above for specific network models.

\section{Numerical examples}
\label{sec:numerics}

To check whether the waveform~\eqref{eq:opt_gen} provides an entrainment of a specific network with a minimum value of the functional~\eqref{eq:func}, we use a trial  function  consisting of two rectangular positive and negative pulses~\cite{Pyragas2018}   
\begin{equation}
	I(\vartheta) = a\left[\Pi\qty( \frac{s\vartheta}{l} ) -\frac{1}{s}\Pi\qty( \frac{\vartheta +d}{l})\right]
	\label{eq:trial}
\end{equation}
separated by a distance $d \in [-\pi, \pi]$. Here $l$ is the width of the negative pulse and $a$ is the amplitude of the positive pulse. The parameter $s$ defines the asymmetry of the pulses. The width of the positive pulse is $l/s$ and the amplitude of the negative pulse is $a/s$, so that the charge balance condition~\eqref{eq:charge_balance} is satisfied. The trial function~\eqref{eq:trial} coincides  with the optimal waveform~\eqref{eq:opt_gen}, $I(\vartheta)=I^*(\vartheta)$, provided that its parameters accord with the optimal values defined by Eqs.~\eqref{eq:delta_theta_z} and \eqref{eq:widths}, i.e., $l=\Delta \vartheta_-$, $a=I_+$ and $d=\Delta \vartheta_{z}$ for $\Delta \omega >0$ and $d=-\Delta \vartheta_{z}$ for $\Delta \omega <0$. For any fixed values of the the parameters $l$ and $d$ of the trial function, we can compute the threshold value of the amplitude $a=a_{th}$ at which the entrainment of the oscillations occurs. The value of the functional \eqref{eq:func} at the entrainment condition is: 
\begin{equation}
	\mathcal{J}_{th}=a_{th}l/s\pi.
	\label{eq:Jc}
\end{equation}
Thus we can verify whether the value of $J_{th}$ is minimal when the parameters  $l$ and $d$ of the trial function coincide with the  optimal values $l=\Delta\vartheta_-$ and $d=\pm \Delta \vartheta_z$. 

Below we estimate the threshold amplitude $a_{th}$ using two different methods: (i) by directly integrating the nonlinear system \eqref{eq:gen_neuron} and (ii) using the approximate phase Eq.~\eqref{eq:phase1}. The first method is straightforward. For given fixed values of the parameters $s$, $l$ and $d$, we vary the amplitude $a$ of the trial signal~\eqref{eq:trial} and find the minimum $a$ at which the entrainment in Eqs.~\eqref{eq:gen_neuron} appears. We interpret this minimum value of $a$ as a threshold amplitude $a_{th}$. In the second method, we introduce the phase difference $\varphi=\vartheta-\omega t$ and, using 
 Eq.~\eqref{eq:phase1}, we derive  the averaged equation for this variable:
\begin{equation}
	\dot{\varphi}=-\Delta \omega  +\frac{1}{2\pi}\int_{-\pi}^\pi z(\vartheta+\varphi) I(\vartheta)d \vartheta.
	\label{eq:aver_phase}
\end{equation}
Then we define an auxiliary function 
%
%
%
\begin{eqnarray}
\Phi(\varphi|s,l,d) = \frac{1}{2\pi a}\int_{-\pi}^\pi z(\vartheta+\varphi) I(\vartheta)d \vartheta 
 \label{eq:auxil_fun}
\end{eqnarray}	
and estimate the threshold amplitude as
\begin{equation}
	a_{th} = 
	\Delta\omega\begin{cases}
1/\displaystyle \max_{\varphi}[\Phi(\varphi|s,l,d)] &\text{for} \quad \Delta \omega >0\\*
1/\displaystyle \min_{\varphi}[\Phi(\varphi|s,l,d)] &\text{for} \quad \Delta \omega <0.
	\end{cases}
	\label{eq:ath}
\end{equation}
We see that the threshold amplitude $a_{th}$ is proportional to the frequency detuning $\Delta \omega$, and therefore $J_{th}$ is also proportional to $\Delta \omega$. The natural characteristic for analysis is $J_{th}/|\Delta \omega|$. We fix the parameters $s$ and $l$ and analyze the dependence of this characteristic on the distance $d$ between the positive and negative pulses.  We  check if $\mathcal{J}_{th}/|\Delta \omega|$ reaches an absolute minimum at $d=\Delta\vartheta_z$ ($d=-\Delta\vartheta_z$) for $\Delta \omega >0$ ($\Delta \omega <0$). We also check whether the values of these minima coincide with the theoretical optimal value $\mathcal{J}^*/|\Delta \omega|$  defined by the Eq.~\eqref{eq:optQs}. We emphasize that the optimal value depends only on the amplitude of the PRC, $\mathcal{J}^{*}/|\Delta \omega|=2/(z_{max}-z_{min})$, and does not depend on any parameters of the control algorithm, such as admissible limits $I_-$ and $I_+$ of the stimulating current.   

Now we present the results of the above analysis for two small-scale FitzHugh-Nagumo neuron networks with synaptic and electric couplings, as well as a large-scale network of synatypically coupled quadratic integrate-and-fire neurons.

\subsection{A network of synaptically coupled FHN neurons }
\label{sec:FHNsyn}

\begin{figure}
\centering
	\includegraphics{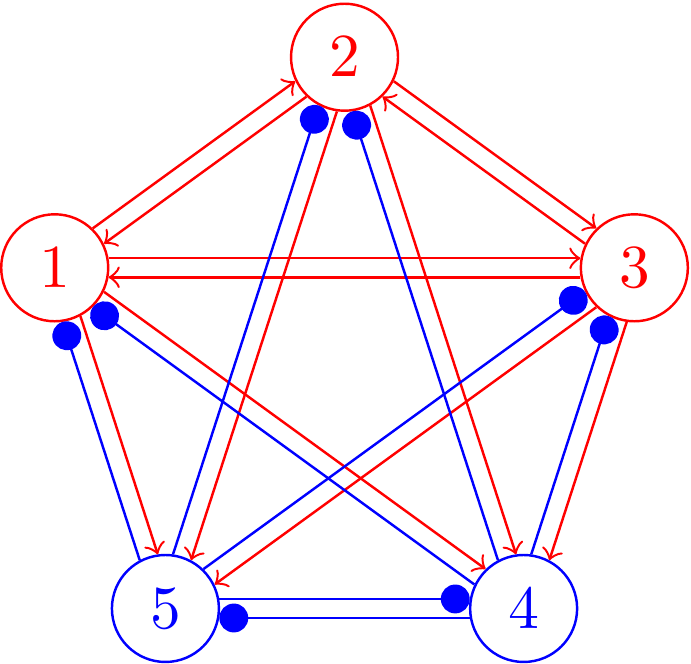}
\caption{\label{network} Schematic diagram of a network of synaptically coupled five FHN neurons. Red circles with numbers 1,2 and 3 are excitatory neurons and blue circles with numbers 4 and 5 are inhibitory neurons.  Red lines ending with arrows indicate excitatory couplings and blue lines ending with filled circles indicate inhibitory couplings. Excitatory neurons are oscillating ($\gamma_i=0.8$ for $i=1,2,3$), and inhibitory neurons are excitable ( $\gamma_i=0.2$ for $i=4,5$).  
}
\end{figure}

First, we verify the theoretical results presented in Sec.~\ref{sec:Optimaltheory} for a network of synaptically coupled FitzHugh-Nagumo neurons. The recovery variable $w_i$ of the FHN neuron is a scalar variable, and the functions on the right side of the Eq.~\eqref{eq:gen_neuron} are as follows
\begin{subequations}
\label{eq:functions}
\begin{eqnarray}
F_i(v_i,w_i) &=& v_i-v_i^3/3-w_i+\gamma_i,  \label{eq:functionsa}\\
G_i(v_i,w_i) &=& \delta (\alpha+v_i-\beta w_i),\label{eq:functionsb} \\
 H_{ij}(v_i,v_j) &=& K_{ij} S_j(v_j). \label{eq:functionsc}
\end{eqnarray}
\end{subequations}
The schematic diagram of the network considered here is shown in Fig.~\ref{network}. It consists of $N=5$ neurons with identical parameters $\alpha=0.7$, $\beta=0.8$, and $\delta=0.08$ for all neurons. The parameters $\gamma_i$, which determine the type of isolated neurons, are not identical. Depending on this parameter, the neuron may be oscillatory or excitable. The values of these parameters are $\gamma_i=0.8$ for the first three neurons, $i=1,2,3$, which exhibit self-oscillatory dynamics, and $\gamma_i=0.2$ for the last two neurons, $i=4,5$, which demonstrate excitable dynamics. 

\begin{figure}
\centering
	\includegraphics{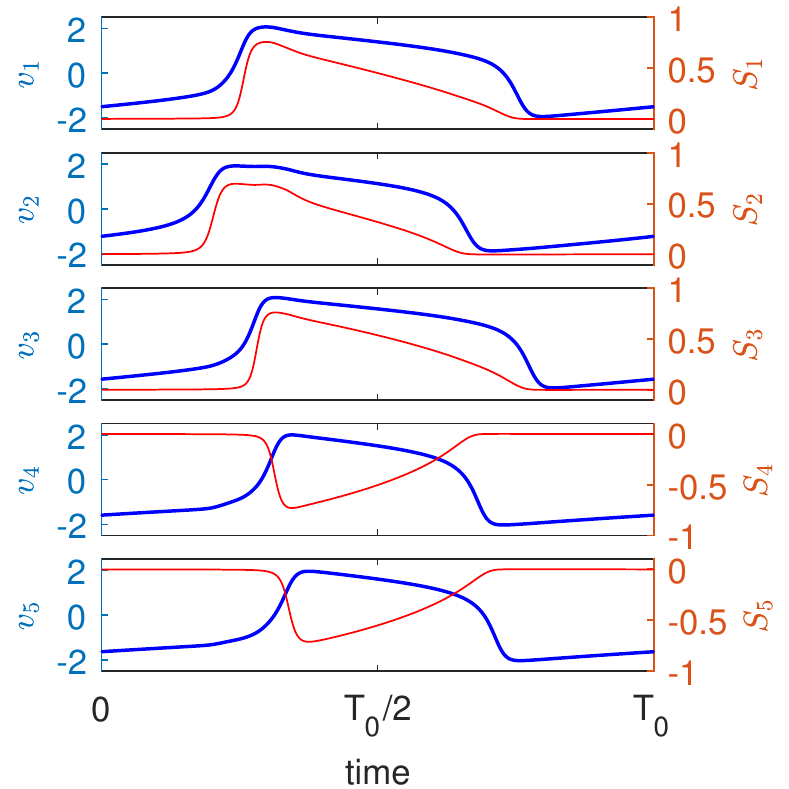}
\caption{\label{FHNsyn1} Dynamics of membrane potentials $v_j$ (bold blue curves) and synaptic variables $S_j$ (thin red curves) of a network of synaptically coupled FHN neurons in the absence of stimulation. 
}
\end{figure}

\begin{figure}
\centering
	\includegraphics{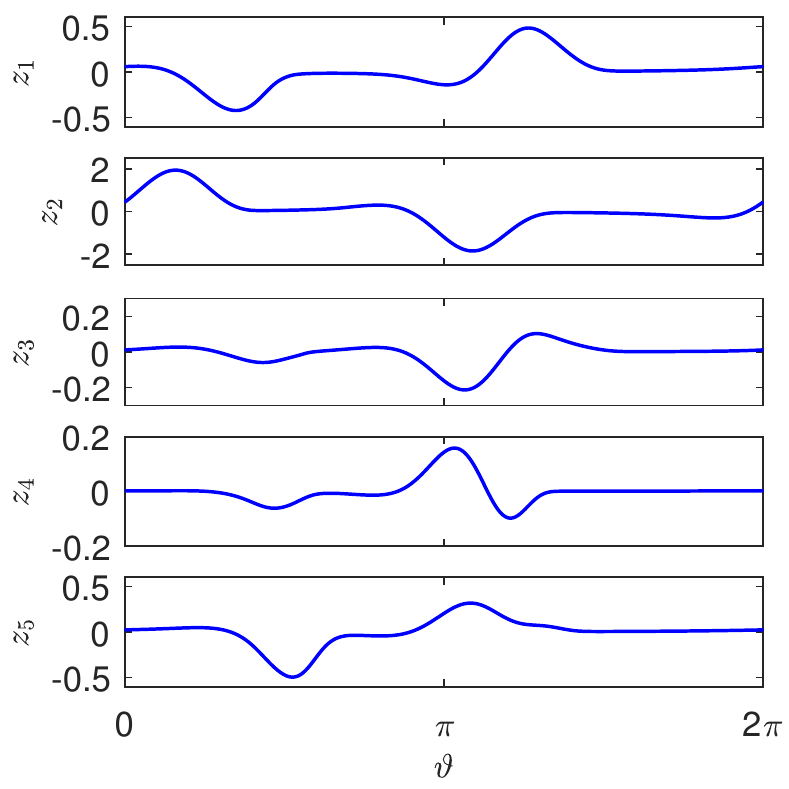}
\caption{\label{FHNsyn2} Phase response curves of a network of synaptically coupled  FHN neurons.
}
\end{figure}

We assume that synaptic dynamics are fast and the synaptic current induced by the $j$th presynaptic neuron to the $i$th postsynaptic neuron can be written by 
Eq.~\eqref{eq:functionsc}, where   $K_{ij}$ is the coupling strength between the $j$th and $i$th neuron,  and $S_j(v_j)$ describes the synaptic spike generated by the $j$th presynaptic neuron. We simulate the dependence of this spike on the voltage $v_j$ of the presynaptic neuron using the sigmoid function
\begin{equation}
S_j(v_j)=p_j \left[1+\exp\left(-\frac{v_j-v_{th}}{\sigma}\right)\right]^{-1},
	\label{eq:sigmoid}
\end{equation}
where $v_{th}$ and $\sigma$ are the characteristic parameters of the synapse and the parameter $p_j$ determines the sign of the synaptic spike: $p_j=+1$ for excitatory neurons and $p_j=-1$ for inhibitory neurons. We assume that the oscillating neurons are excitatory ($p_j=+1$ for $j=1,2,3$) and excitable neurons are inhibitory ($p_j=-1$ for $j=4,5$). Synaptic parameters are $v_{th}=1.5$ and $\sigma=0.5$. Elements $K_{ij}$ of the coupling matrix are randomly and independently taken from a uniform distribution $[0,0.5]$. 
Below we present the results for a specific realization of a randomly generated matrix:
\begin{equation}
K = 
\begin{pmatrix}
         0   & 0.4710  &  0.2769  &  0.4334  &  0.2007\\
    0.0855   &      0  &  0.3400  &  0.2034  &  0.4167\\
    0.4693   & 0.2260  &       0  &  0.0563  &  0.2018\\
    0.2952   & 0.4198  &  0.1196  &       0  &  0.1951\\
    0.2203   & 0.2663  &  0.2895  &  0.1501  &       0
\end{pmatrix}.
\label{eq:coupl_matrix}
\end{equation}
The zero diagonal elements mean that self-coupling is excluded. For given parameter values, the free network demonstrates collective limit-cycle oscillations with a period $T_0 \approx 35.159894$. The dynamics of the membrane potentials $v_j(t)$ and the spike variables $S_j(v_j(t))$ for all neurons are shown in Fig.~\ref{FHNsyn1}. Figure~\ref{FHNsyn2} shows the corresponding PRCs $z_j(\vartheta)$ obtained by solving the Eqs.~\eqref{eq:adjoint}. 

The network architecture shown in Fig.~\ref{network}, mimics the architecture of the neural network of the subthalamic nucleus (STN) and the external segment of the globus pallidus (GPe), which is often used to model Parkinson's disease (cf.,e.g., Ref.~\cite{Terman2002}). STN is a network of oscillating excitatory neurons (in our case, the first three neurons), and GPe consists of excitable inhibitory neurons (in our case, the last two neurons). Using the calculated PRCs, we can build the optimal waveform for various stimulation protocols. 

\begin{figure}
\centering
	\includegraphics{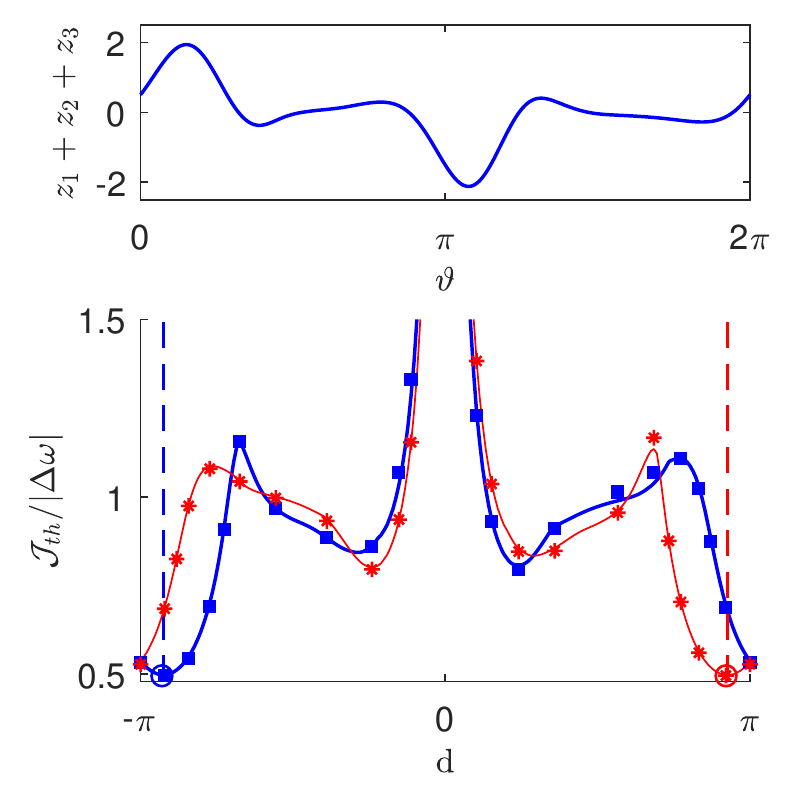}
\caption{\label{FHNsyn3} Testing the minimum-charge control theory for a network of five synaptically coupled FHN neurons in the case of stimulation of only oscillating excitatory neurons with the indexes $j=1,2,3$. (Top) Effective PRC $z=z_1+z_2+z_3$. The phase distance between the maximum and the minimum is  $\Delta \vartheta_z=-2.9084$, and the amplitude of the PRC is $z_{max}-z_{min}=4.0634$. (Bottom) Threshold value $\mathcal{J}_{th}$ of the functional \eqref{eq:func}, normalized to the frequency detuning $|\Delta \omega|$, as a function of the distance $d$ between positive and negative pulses of the trial waveform~\eqref{eq:trial} for the fixed parameters $l=0.2$ and $s=2$. The bold blue curve and the the thin red curve show the results obtained from the averaged phase Eq.~\eqref{eq:aver_phase} for  $\Delta \omega>0$ and  $\Delta \omega<0$, receptively. Blue squires and red stars denote the corresponding results obtained by integrating a system of nonlinear Eqs.~\eqref{eq:gen_neuron}. The dashed vertical lines show the theoretical optimal distances $d=\pm \Delta \vartheta_z$. Open circles indicate the optimal theoretical value~\eqref{eq:optQs}. 
} 
\end{figure}

\begin{figure}
\centering
	\includegraphics{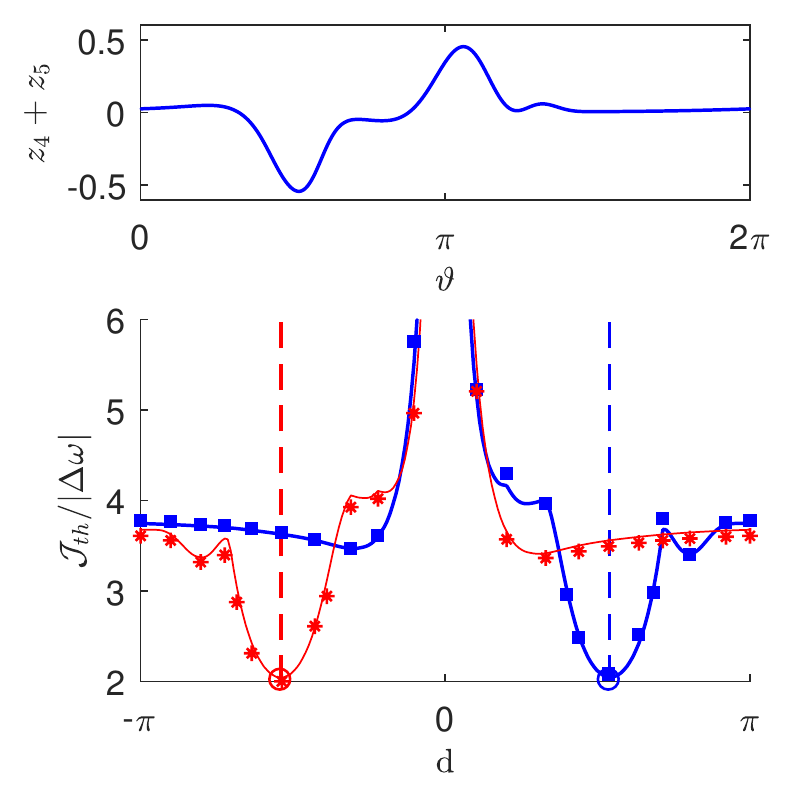}
\caption{\label{FHNsyn4} The same as in Fig.~\ref{FHNsyn4}, but for the case of stimulation of only excitable inhibitory neurons with the indexes $j=4,5$. The effective PRC is $z=z_4+z_5$. The phase distance between the maximum and the minimum is $\Delta \vartheta_z=1.6935$, and the amplitude of the PRC is $z_{max}-z_{min}=0.9949$. The remaining parameters are the same as in  Fig.~\ref{FHNsyn3}.
}
\end{figure}

Below we consider two options for stimulation: (i) only oscillating excitatory neurons with indexes $j = 1,2,3$ are stimulated. Here the effective PRC is $z = z_1 + z_2 + z_3$; (ii) only excitable inhibitory neurons with indexes $j = 4,5$ are stimulated. Here the effective PRC is $z = z_4 + z_5$. The corresponding effective PRC and dependence of $\mathcal{J}_{th}/|\Delta \omega|$ on the distance $d$ between the pulses of the trial waveform~\eqref{eq:trial} for  (i) and (ii) stimulation protocols are shown in Fig.~(\ref{FHNsyn3}) and (\ref{FHNsyn4}), respectively. Other parameters of the trial waveform are fixed, $l=0.2$ and $s=2$. The solid blue curves show the results obtained by the averaged phase Eq.~\eqref{eq:aver_phase} for 
positive frequency detuning $\Delta \omega>0$, and the thin red curves show the same results for negative frequency detuning $\Delta \omega<0$. Symbols denote the results obtained by integrating a system of nonlinear Eqs.~\eqref{eq:gen_neuron}. Theoretical optimal distances $d=\pm \Delta \vartheta_z$ between positive and negative pulses are shown by vertical dashed lines. Open circles indicate the theoretical optimal  value~\eqref{eq:optQs} of the functional~\eqref{eq:func}, normalized to the frequency detuning $|\Delta \omega|$. 

In both cases, the solution of the averaged phase Eq.~\eqref{eq:aver_phase} and the direct simulation of a nonlinear system of Eqs.~\eqref{eq:gen_neuron} confirm the theoretical results presented in the Sec.~\ref{sec:Optimaltheory}. For $\Delta \omega >0$, the absolute minimum of the functional~\eqref{eq:func} is attained when the distance $d$ between the positive and negative pulses of the trial waveform~\eqref{eq:trial} coincides with distance $\Delta \vartheta_z$ between the absolute maximum and the absolute minimum of the corresponding PRC. For $\Delta \omega <0$, the absolute minimum is attained at $d=-\Delta \vartheta_z$. The values of these minima  are in good agreement with the theoretically predicted value~\eqref{eq:optQs}. 
Also note the good agreement between the results obtained from the averaged phase Eq~\eqref{eq:aver_phase} and the nonliear system~\eqref{eq:gen_neuron} when calculating the dependence $\mathcal{J}_{th}/|\Delta \omega|$ vs. $d$ in the entire interval $d \in [-\pi, \pi]$.

From a physical point of view, an interesting result is that the first stimulation protocol is more effective than the second. For the first protocol, the total charge delivered to each neuron during the stimulation period is four times less. This is due to the fact that oscillating excitatory neurons are more sensitive to  external perturbations than excitable inhibitory neurons. It can be seen from the effective PRCs of these two subsystems shown in the Figs.~\ref{FHNsyn3} and \ref{FHNsyn4}. The amplitudes of these PRCs differ four times. The optimal value of the functional~\eqref{eq:optQs} is inversely proportional to the amplitude of the PRC and, therefore, for the first stimulation protocol is four times less than for the second. We also note that the optimal distance between the positive and negative pulses of the stimulation current is different for these two stimulation protocols, because the distance between the maximum and the minimum of the corresponding effective PRCs is different.

\subsection{A network of electrically coupled FHN neurons}
\label{sec:FHNdif}

As a second example, we consider the network of electrically coupled FHN neurons introduced in Ref.~\cite{Nakao2018}. The network size $N=10$ is two times larger than in the previous example. As before, the functions $F_i(v_i,w_i)$ and $G_i(v_i,w_i)$ are defined by the Eqs.~\eqref{eq:functionsa} and \eqref{eq:functionsb}, respectively, and the function $H_{ij}(v_i,v_j)$ is now described by an electric coupling of the form  
\begin{equation}
H_{ij}(v_i,v_j)=K_{ij}(v_j-v_i). 
	\label{eq:electr_coupl}
\end{equation}
As in the previous example, the parameters $\alpha=0.7$, $\beta=0.8$, and $\delta=0.08$ are the same for all neurons, and the parameters $\gamma_i$ are not identical. The values of these parameters are $\gamma_i=0.2$ for the neurons $i=1,\ldots, 7$, which exhibit excitable dynamics, and $\gamma_i=0.8$ for the neurons $i=8,\ldots, 10$, which demonstrate self-oscillatory dynamics. The elements $K_{ij}$ of the $10\times 10$ coupling matrix are generated randomly. Here we use the specific realization of this matrix presented in Ref.~\cite{Nakao2018}.  For the above parameter values, the free $[I_i(\omega t)=0]$ network demonstrates collective limit-cycle oscillations in the $20$-dimensional state space with the period $T_0 \approx 75.709874$. The dynamics of the variables $v_i(t)$ and the PRCs $z_i(\vartheta)$ for all neurons are graphically presented in Ref.~\cite{Nakao2018}. 

\begin{figure}
\centering
	\includegraphics{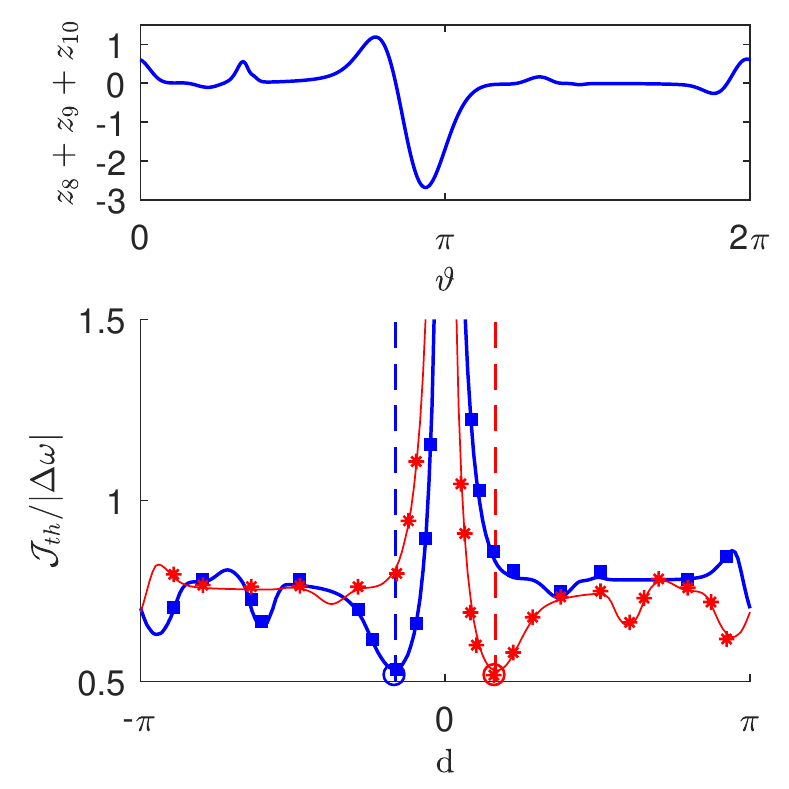}
\caption{\label{FHNdiff1} Testing the minimum-charge control theory for a network of ten electrically coupled FHN neurons in the case of stimulation of only oscillating neurons with the indexes $j=8,9,10$.
(Top) Effective PRC $z=z_8+z_9+z_{10}$. The phase distance between the maximum and the minimum is  $\Delta \vartheta_z=-0.5154$, and the amplitude of the PRC is $z_{max}-z_{min}=3.8814$. (Bottom) The dependence  $\mathcal{J}_{th}/|\Delta \omega|$ vs. $d$ for the fixed parameters $l=0.2$ and $s=2$. Marking of curves and symbols is the same as in Fig.~\ref{FHNsyn3}.
}
\end{figure}

In Fig.~\ref{FHNdiff1}, we show the results for the stimulation  protocol, when only oscillating neurons with indexes $i=8,9,10$ are stimulated. The effective PRC of the network, $z=z_8+z_9+z_{10}$ (the upper graph) is now more complex  then in previous examples; it has more extrema. This leads to a more complex dependence of $\mathcal{J}_{th}/|\Delta \omega|$ on $d$ (the lower graph), which also has more extrema. However, as in  previous examples, the absolute minimum of this dependence is located at $d=\vartheta_z$ for $\Delta \omega <0$ and at $d=-\vartheta_z$ for $\Delta \omega <0$, where $\vartheta_z  \approx -0.5154$ is the distance between the absolute maximum and the absolute minimum of the PRC. The absolute minimum value of $\mathcal{J}_{th}/|\Delta \omega|$  is consistent with the Eq.~\eqref{eq:optQs}. Therefore, the more complex network model discussed here also confirms the general theory presented in the Sec.~\ref{sec:Optimaltheory}. 

\subsection{A large-scale network of synatypically coupled quadratic integrate-and-fire neurons}
\label{sec:syn}

As a final example, we consider a heterogeneous network with a large number $N$ of all-to-all synaptically coupled QIF neurons. The microscopic state of the network is determined by the set of neurons' membrane potentials $v_i$, which satisfy the following equations~\cite{ermentrout10}:
\begin{equation}
\dot{v}_{i}=v_{i}^{2}+\eta_{i}+S(t)+I(\omega t) \quad (i=1,\ldots, N). \label{Vj}
\end{equation}
Here, the  constants $\eta_i$ specify the behavior of individual neurons, $S(t)$ denotes the synaptic current and the last term $I(\omega t)$ represents a periodic stimulating current. We assume that all neurons are stimulated by the same external signal. The QIF neuron model does not contain a recovery variable. Recovery is described by an instantaneous reset of the membrane potential. Every moment when the membrane potential $v_i$ reaches the peak value $v_{peak}$ its voltage is reset to the value  $v_{reset}$. To simplify the analysis, we set $v_{peak}=-v_{reset} \to \infty$. We also assume that the synaptic dynamics is fast and synaptic current can be written as~\citep{Ratas2016}:
\begin{equation}
S(t)=J\frac{v_{th}}{N}\sum_{j=1}^{N}H(v_{j}(t)-v_{th}). \label{coupl_def}
\end{equation}
Here $J$ represents the coupling strength, $H(\cdot)$ is the Heaviside step function,  and $v_{th}$ is the threshold potential. The positive and negative signs of $J$ correspond to the excitatory and inhibitory interactions, respectively. At 
time $t$, only those neurons contribute to the synaptic current, whose membrane potential $v_j(t)$ exceeds the threshold value $v_ {th}$. In fact, the $v_ {th}$ parameter determines the width and height of the synaptic pulses. When the $j$th neuron spikes, the therm $v_{th} H(v_{j}(t)-v_{th})$ generates a rectangular pulse of height $v_{th}$. The pulse width  for large $v_{th}$ can be approximated  as $1/v_{th}$~\cite{Ratas2016}. When $v_{th}\to\infty$, the pulse turns into a zero-width Dirac delta spike. The case of interaction with instantaneous Dirac delta pulses was considered in Ref.~\cite{Montbrio2015}, and macroscopic limit cycle oscillations were not found in such a model. The finite width of synaptic pulses is a crucial factor for the occurrence of macroscopic self-sustained oscillations~\cite{Ratas2016}.

The isolated [$S(t)=0, I(\omega t)=0$] QIF neuron is the canonical model for the class I neurons near the spiking threshold~\cite{izhi07}. 
Spiking instability in such neurons is manifested through bifurcation of the saddle node on the invariant curve (SNIC). The system following this scenario exhibits excitability before the bifurcation. For the QIF neuron, this scenario is provided by the bifurcation parameter $\eta_i$. For $\eta_i<0$, the neuron is in the excitable mode and for $\eta_i>0$ it is in the spiking mode. We assume that the values of the parameters $\eta_i$ are distributed in accordance with 
a bell-shaped probability density function,  which can be approximated by the Lorentzian distribution:
\begin{equation}
g(\eta)=\frac{1}{\pi} \frac{\Delta}{(\eta-\bar{\eta})^2+\Delta^2}, \label{Lor}
\end{equation}
where $\Delta$ and $\bar{\eta}$ are the width and the center of the distribution, respectively. 

The advantage of this model is that it allows an exact low-dimensional reduction of system equations in the thermodynamic limit of infinite number of neurons, $N \to \infty$. In this limit, one can derive the closed system of two ordinary differential equations for biophysically relevant macroscopic quantities, the mean membrane potential $v(t)$ and the firing rate $r(t)$~\cite{Ratas2016}:
\begin{subequations}
\label{eq_rv1}
\begin{eqnarray}
\dot{v} & = & \bar{\eta} +v^2-\pi^2 r^2+S(v,r)+I(\omega t),\label{eq_v1}\\
\dot{r} & = & \Delta/\pi+ 2rv\label{eq_r1}
\end{eqnarray}
\end{subequations}
Here, the synaptic current $S=S(v,r)$ is a function of the variables $v$  and $r$ of the following form:
\begin{equation}
S(v,r)=J\frac{v_{th}}{\pi} \left[\frac{\pi}{2}- \arctan \left(  \frac{v_{th}-v}{ \pi r }  \right)\right]. \label{St2}
\end{equation}
The low-dimensional macroscopic model greatly simplifies the task of finding an effective PRC  for the original  microscopic model determined by a large system of Eqs.~\eqref{Vj}. For large $N$, the macroscopic model \eqref{eq_rv1} approximates well the solutions of the microscopic model~\eqref{Vj}, and therefore the PRC for the microscopic model can be obtained from the above system of Eqs.~\eqref{eq_rv1}. Consider the case when the free [$I(\omega t)=0$] system~\eqref{eq_rv1} has a limit cycle solution $[v^{(0)}(t), r^{(0)}(t)]= [v^{(0)}(t+T_0), r^{0}(t+T_0)]$ with the period $T_0$. Then the PRC $\mathbf{Q}=[Q^{(1)}, Q^{(2)}]^T$ of the reduced system \eqref{eq_rv1} satisfies the adjoint equation:
\begin{equation} \label{eq:adjoint1}
\omega_0 \frac{d}{d \vartheta}\mathbf{Q}(\vartheta)= -A^T(\vartheta)\mathbf{Q}(\vartheta),
\end{equation}
where $\omega_0=2\pi/T_0$ and 
\begin{equation}
A(\vartheta)= 
\begin{pmatrix}
\partial S(v,r)/\partial v +2v & \partial S(v,r)/\partial r -2\pi^2 r \\
    2r  &       2v
\end{pmatrix}
\label{eq:Jacob}
\end{equation}
is the Jacobian matrix of the system \eqref{eq_rv1} evaluated at $[v^{(0)}(\vartheta), r^{(0)}(\vartheta)]= [v^{(0)}(\omega_0 t), r^{0}(\omega_0 t)]$. 

To summarize, solving the adjoint Eq.~\eqref{eq:adjoint1}, we can find the PRC $\mathbf{Q}(\vartheta)$ of the reduced system \eqref{eq_rv1}. The first component of this PRC $z(\vartheta)=Q^{(1)}(\vartheta)$ can be used to describe the phase dynamics of the original large-scale system~\eqref{Vj} in the presence of a weak stimulation current $I(\omega t)$. 
This dynamics is described by the phase Eq.~\eqref{eq:phase1}, which is the basis for the optimal theory presented in the Sec.~\ref{sec:Optimaltheory}. Thus, the results of this theory are applicable to a large-scale network~\eqref{Vj} of QIF neurons with an effective PRC $z(\vartheta)$ defined by a simple adjoint Eq.~\eqref{eq:adjoint1}. Below we support this statement with a specific numerical example.

\begin{figure}
\centering
	\includegraphics{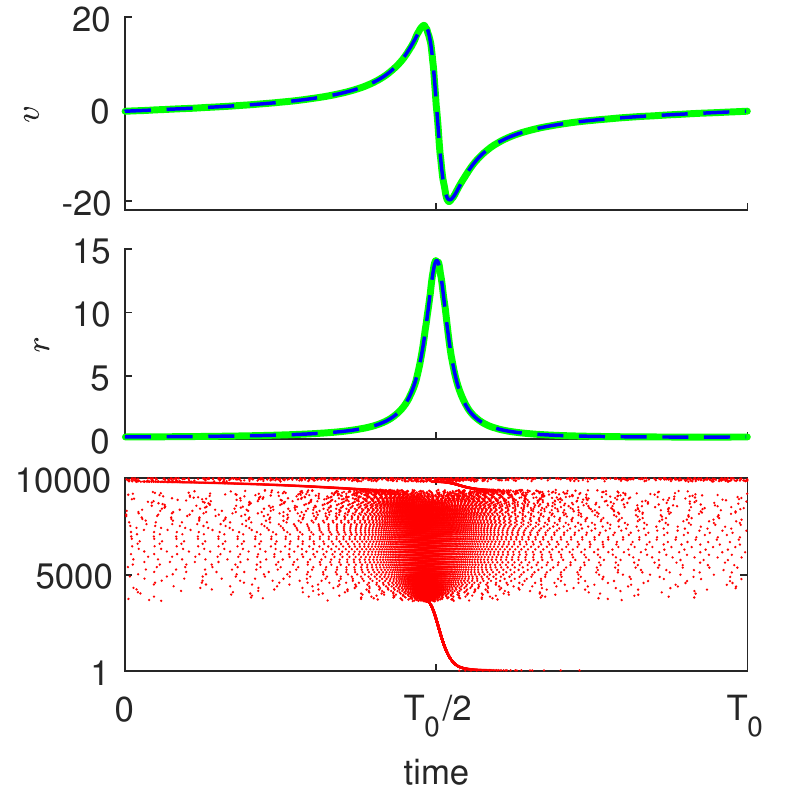}
\caption{\label{QIF1} Dynamics of a heterogeneous network of $10^4$ synapticlly coupled QIF neurons in the absence of stimulation for parameter values $v_{th}=50$, $J=30$, $\Delta=1$ and $\bar{\eta}=0$. The upper and middle graph show the evolution of the mean membrane potential $v(t)$ and the spiking rate $r(t)$, respectively. The thin dashed blue curves show the solutions of the reduced system of Eqs.~\eqref{eq_rv1}, and the bold solid green curves show the results of direct numerical simulation of a microscopic model of Eqs.~\eqref{theta_j} for $N=10^4$ neurons. The bottom graph shows the raster plot. Here the dots show the spike moments for each neuron, where the vertical axis indicates neuron numbers.  
}
\end{figure}

We consider the network of $N=10^4$ QIF neurons with parameter values $v_{th}=50$, $J=30$, $\Delta=1$ and $\bar{\eta}=0$. Choosing a zero value for the parameter $\bar{\eta}$ means that half of the neurons in the network are oscillating and the other half are excitable. For these parameter values, the macroscopic model~\eqref{eq_rv1} shows limit cycle oscillations with a period of $T_0 \approx 1.130132$. The dynamics of the mean membrane potential $v(t)$ and the spiking rate $r(t)$ during one oscillation period are shown by thin dashed blue curves in the upper and middle graphs, respectively, in Fig.~\ref{QIF1}.

Numerical simulation of the microscopic model~\eqref{Vj} is more convenient after changing the variables
\begin{equation}
v_i = \tan(\theta_i/2)\label{eq_transf}
\end{equation}
that turn QIF neurons into theta neurons. Such a transformation of variables avoids the problem associated with jumps of infinite size (from $+ \infty $ to $-\infty $) of the membrane potential $v_i$ of the QIF neuron at the moments of firing. The phase $\theta_i$ of the theta neuron simply crosses the value of $\theta_i=\pi$ at these moments. For theta neurons, the Eqs.~\eqref{Vj} are transformed into
\begin{equation}
\dot{\theta}_{i}= 1-\cos \theta_{i}+(1+\cos \theta_{i})\left[\eta_{i}+S(t)+I(\omega t) \right], \label{theta_j}
\end{equation}
where the synaptic current $S(t)$ is determined by the Eqs.~\eqref{coupl_def} and \eqref{eq_transf}. These equations were integrated by the Euler method with a time step of $d t = 10^{-4}$. The population of $N=10^4$ theta neurons with the Lorentzian distribution \eqref{Lor} were deterministically generated using $\eta_j=\bar{\eta}+\Delta \tan\left[(\pi/2)(2j-N-1)/(N+1)\right]$,  where $j=1,\ldots, N$, $\Delta=1$ and $\bar{\eta}=0$. Such a numbering of  neurons means that free neurons with the indexes $j=1,\ldots, 5000$ are excitable and neurons with the indexes $j=5001,\ldots, 10000$ are oscillating. More information on numerical modeling of Eqs.~\eqref{theta_j} can be found in Ref.~\cite{Ratas2016}. To compare the results obtained from the microscopic model~\eqref{theta_j} with the solutions of the reduced system~\eqref{eq_rv1}, we calculate the Kuramoto order parameter~\cite{Kuramoto2003}
\begin{equation}
\label{eq_Z}
Z=\frac{1}{N}\sum\limits_{j=1}^{N}\exp(i \theta_j)
\end{equation} 
and use the relationship between $Z$ and the macroscopic parameters $v$ and $r$~\cite{Montbrio2015}: 
\begin{equation}
\label{eq_W}
v=\operatorname{Im}\left(\frac{1-Z^*}{1+Z^*}\right), \quad r=\frac{1}{\pi}\operatorname{Re}\left(\frac{1-Z^*}{1+Z^*}\right),
\end{equation} 
where $Z^*$ means complex conjugate of $Z$. In Fig.~\ref{QIF1}, the dynamics of the mean membrane potential $v(t)$ and the spiking rate $r(t)$ estimated from the microscopic model of Eqs.~\eqref{theta_j}, \eqref{eq_Z} and \eqref{eq_W} are shown by bold solid green curves in the upper and the middle graphs, respectively. These solutions are in excellent agreement with the solutions of the  macroscopic model~\eqref{eq_rv1}, shown by thin dashed blue curves. Note that, in contrast to the macroscopic model, the variables $v(t)$ and $r(t)$ obtained from the microscopic model are not exactly periodic. Their period fluctuate around a mean value of $T_0 \approx 1.1348$ with a standard deviation of of about half a percent. These fluctuations are related to the finite size of the network. The microscopic dynamics of the network is quite complex. This can be seen from the raster plot shown in the bottom graph, where dots indicate the spike moments of each neuron. 

\begin{figure}
\centering
	\includegraphics{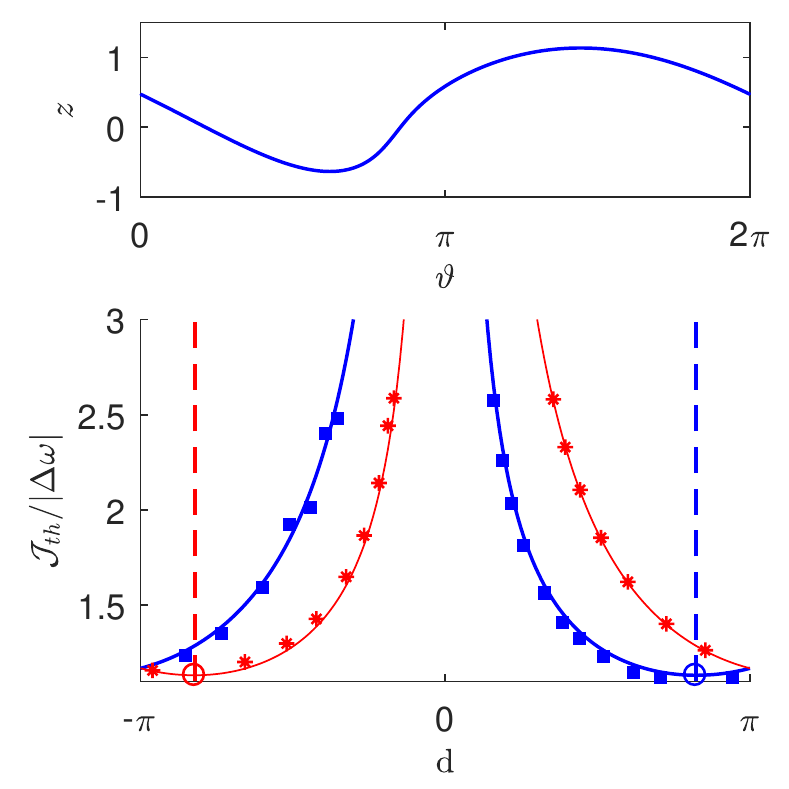}
\caption{\label{QIF2} Testing the minimum-charge control theory for a network of $10^4$ synaptically coupled QIF neurons. (Top) Effective PRC of the network in the case of homogeneous stimulation of all neurons. The PRC is obtained from a simple adjoint Eq.~\eqref{eq:adjoint1}. The phase distance between the maximum and the minimum is  $\Delta \vartheta_z=2.5832$, and the amplitude of the PRC is $z_{max}-z_{min}=1.7696$. (Bottom) Dependence  $\mathcal{J}_{th}/|\Delta \omega|$ vs. $d$ for fixed parameters $l=0.2$ and $s=2$. Bold blue and  thin red curves are obtained in the same way as in Fig.~\ref{FHNsyn3}, but using the PRC shown in the top graph of this figure. Symbols are obtained by solving a system of $10^4$ microscopic Eqs.~\eqref{theta_j} with a trial stimulation current~\eqref{eq:trial}.
}
\end{figure}

Despite the complex microscopic dynamics of the network, its macroscopic behavior is well described by the reduced system of Eqs.~\eqref{eq_rv1}. The effective PRC of the network obtained from the simple adjoint Eq.~\eqref{eq:adjoint1} is shown in the upper graph in Fig.~\ref{QIF2}. The PRC parameters needed to design  the optimal waveform are $\Delta \vartheta_z=2.5832$ and $z_{max}-z_{min}=1.7696$. As in the previous examples, the bottom graph show the dependence $\mathcal{J}_{th}/|\Delta \omega|$ from the distance $d$ between the positive and negative pulses of the trial waveform ~\eqref{eq:trial} for the fixed  parameters $l=0.2$ and $s=2$. Bold blue and  thin red curves are obtained from the phase Eq.~\eqref{eq:phase1} for $\Delta\omega>0$ and $\Delta\omega<0$, respectively. Symbols show the results obtained from the microscopic model. Threshold entrainment amplitudes $a_{th}$ were found by solving the system of $10^4$ Eqs.~\eqref{theta_j} with a trial signal~\eqref{eq:trial}. We see that both results are in good agreement with each other. They  show that the minimum of $\mathcal{J}_{th}/|\Delta \omega|$ is attained at $d=\vartheta_z$ for $\Delta \omega >0$ and at $d=-\vartheta_z$ for $\Delta \omega <0$, where $\vartheta_z  \approx 2.5832$ is the distance between the maximum and the minimum of the PRC. The minimum value of $\mathcal{J}_{th}/|\Delta \omega|$  is consistent with the theoretically predicted optimal value~\eqref{eq:optQs}. Thus, the minimum charge control theory presented in Sec.~\ref{sec:Optimaltheory} works well not only for small neural networks, but also for a large-scale network of interacting QIF neurons, the collective behavior of which exhibits periodic macroscopic oscillations.

\section{Discussion}
\label{sec:conclusions}

We examined the problem of optimal entrainment of a network of interacting neurons by an external stimulus when an unperturbed network exhibits collective periodic oscillations. The general expression for the optimal waveform, which provides network entertainment with a minimum mean absolute value of the periodic stimulating current, is presented. This optimization is clinically relevant because it aims to reduce damage to nerve tissue by minimizing the integral charge transferred to neurons in both directions during the stimulation period.
Our optimal waveform satisfies the clinically mandatory requirements: the  charge-balance condition and the amplitude limitation.

The research presented in this paper is based on our recent publication \cite{Pyragas2018}, where we considered a similar problem for the case of a single neuron. We obtained the optimal waveform under the assumption of a small frequency detuning, when the equations of the neuron model can be reduced to a simple scalar equation for the phase.
Here we showed that, under certain assumptions, network equations can also be reduced to the same phase equation as for a single neuron. The only difference is that the phase response curve of a single neuron is replaced by an effective phase response curve of the network. This allowed us to adapt the results of the minimum charge control theory developed in Ref.~\cite{Pyragas2018}. As well as for a single neuron, the optimal waveform of the network is of bang-off-bang type with periodically repeated positive and negative rectangular pulses, generally of different amplitudes and widths. The distance $d$ between positive and negative  pulses is determined by the distance $\Delta\vartheta_z$ between the absolute maximum and the absolute minimum of the effective phase response curve. For the positive frequency deturning, the optimal distance between the pulses is $d=\Delta\vartheta_z$, and for the negative frequency deturning, $d=-\Delta\vartheta_z$.

We confirmed the theoretical results with three numerical examples: two small-scale networks  consisting of (i) five synaptically coupled FHN neurons, (ii) ten electrically coupled FHN neurons, and (iii) a large-scale network with $10^4$ synaptically coupled QIF neurons. In the first example, the network architecture  mimics the network architecture of the STN-GPe model~\cite{Terman2002}, which consists of oscillating excitatory (STN) and excitable inhibitory (GPe) neurons. Two stimulation protocols were considered. In the first protocol, only oscillating excitatory neurons were stimulated, and in the second --- only excitable inhibitory neurons. Our results showed that the first stimulation protocol is more effective. For the first protocol, the entrainment of the network by an external stimulus was achieved with a four time lower mean absolute value of the  stimulating current than for the second. The second example demonstrated the validity of our theory for the network of electrically coupled oscillating and excitable FHN neurons introduced in Ref.~\cite{Nakao2018}. Finally, in the third example, we used the QIF neural network model, which allows an exact low-dimensional reduction of system equations in the thermodynamic limit of an infinite number of neurons~\cite{Montbrio2015,Ratas2016}. Based on the reduced macroscopic model, we derived a simple adjoint equation for the phase response curve, which is necessary to construct the optimal waveform for a large-scale network. The validity of the optimal waveform was confirmed by direct numerical simulation of a network consisting of $10^4$ synaptically interacting QIF neurons. Although we presented the results for the case when all neurons of the network are stimulated by the same external current, our approach can be extended to the case of heterogeneous stimulation. In this case, the macroscopic model obtained in the thermodynamic limit will have a higher dimension. 

\bibliography{minimumcharge}

\end{document}